\documentclass[aps,prl,reprint,preprintnumbers,nofootinbib,superscriptaddress]{revtex4-1}
\pdfoutput=1
\usepackage{natbib}
\usepackage{amsmath}
\usepackage{amssymb}
\usepackage{amsfonts}
\usepackage{hyperref}
\usepackage{epsfig}
\usepackage{graphicx}
\usepackage{slashed} 
\usepackage{mathrsfs} 
\usepackage{subcaption}
\allowdisplaybreaks

\def\beq{\begin{equation}}
\def\eeq{\end{equation}}
\def\bea{\begin{eqnarray}}
\def\eea{\end{eqnarray}}
\def\ba{\begin{array}}
\def\ea{\end{array}}

\def\t1{\tilde{t}_1}
\def\t2{\tilde{t}_2}
\def\b1{\tilde{b}_1}

\newcommand{\bd}{\begin{displaymath}}
\newcommand{\ed}{\end{displaymath}}
\newcommand{\be}{\begin{equation}}
\newcommand{\ee}{\end{equation}}

\def\b{\beta}

\def\q2 {q^2}

\def\t {\times }

\def\bt{\begin{table}}
\def\et{\end{table}}

\catcode`@=11 
\def \gsim{\mathrel{\mathpalette\@versim>}}
\def \lsim{\mathrel{\mathpalette\@versim<}}
\def \@versim#1#2{\lower0.4ex\vbox{\baselineskip\z@skip\lineskip\z@skip
     \lineskiplimit\z@\ialign{$\m@th#1\hfil##\hfil$%
     \crcr#2\crcr\sim\crcr}}}
\catcode`@=12 

\begin{document}
\preprint {HRI-RECAPP-2017-016}

\title{Do astrophysical data disfavour the\\
        minimal supersymmetric standard model?}

\author{Arpan Kar}
\thanks{E-mail: arpankar@hri.res.in}
\affiliation{Regional Centre for Accelerator-based Particle
  Physics, Harish-Chandra Research Institute, HBNI, Chhatnag Road,
  Jhunsi, Allahabad - 211 019, India}
\author{Sourav Mitra}
\affiliation{Surendranath
  College, 24/2 M. G. ROAD, Kolkata, West Bengal 700009}
\author{Biswarup Mukhopadhyaya}
\affiliation{Regional Centre for Accelerator-based Particle
  Physics, Harish-Chandra Research Institute, HBNI, Chhatnag Road,
  Jhunsi, Allahabad - 211 019, India}
\author{Tirthankar Roy Choudhury}
\affiliation{National Centre for Radio Astrophysics, TIFR, Post Bag 3, Ganeshkhind, Pune 411007, India}

\begin{abstract}
If the minimal supersymmetric standard model (MSSM) is the only new
physics around the TeV-scale, it has to account for the entire dark matter relic
density, or else it will cease to be `minimal'. We use this
expectation to obtain the best quantitative explanation of the
galactic centre $\gamma$-ray excess. The $\gamma$-ray data/flux limits
from other astrophysical sources are also taken into account, together with all
laboratory constraints. The lower limit on the relic density, together
with the latest direct dark matter search constraints and the shape of
the galactic centre $\gamma$-ray spectrum, makes the MSSM fits appear
rather poor. A comparison with similar fits of the Higgs boson mass from
indirect and direct search results makes one suspicious that the MSSM
is not a good explanation of data related to dark matter.
\end{abstract}
 
\maketitle


{\em Introduction: The minimal supersymmetric standard model (MSSM) is
  a popular theoretical scenario explaining the dark matter (DM)
  content of the universe, with the lightest neutralino ($\chi^0_1$) as
  the `dark' particle. However, it is also implicit in the adoption of
  the MSSM as the description of nature around the TeV scale that this
  neutralino should saturate practically the entire relic density of dark matter;
  otherwise it is tantamount to going beyond the MSSM.  This becomes
  an important input in the analysis of $\gamma$-ray data from the
  galactic centre (GC) \cite{TheFermi-LAT:2015kwa, Calore:2014xka,
    TheFermi-LAT:2017vmf} and other celestial objects
  \cite{Geringer-Sameth:2015lua}. Coupled with laboratory constraints
  and those from direct DM search \cite{Aprile:2017iyp,
    PhysRevLett.118.251302}, it makes
  the MSSM an unfavourable fit. We show this with a comparison with
  the analysis of data related to the Higgs boson, where fits of
  indirect data preceded the direct search constraints
  \cite{Baak:2011ze, Flacher:2008zq}.}

Accelerator data, especially those from the Large Hadron Collider
(LHC), have pushed the lower mass limits on most strongly interacting
superparticles in the MSSM to about 2 TeV \cite{Sirunyan:2017cwe,
  Aaboud:2017bac}. On the other hand, if the excess in GC
$\gamma$-rays (principally in the range $1 - 10$ GeV
\cite{TheFermi-LAT:2015kwa, Calore:2014xka}) have to arise from DM
annihilation, then the nature of the spectrum in the energy band
$\approx 1 - 10$ GeV suggests a relatively light ([$\lsim 80, 205]$
GeV) $\chi^0_1$.  One of course has to satisfy all accelerator limits
constraints from direct DM search \cite{Aprile:2017iyp,
  PhysRevLett.118.251302}, other $\gamma$-ray data from dwarf galaxies
like Reticulum II \cite{Geringer-Sameth:2015lua, Zhao:2017pcz,
  Drlica-Wagner:2015xua}, neighbouring galaxies such as M31
\cite{Ackermann:2017nya}, and upper limits on radio synchrotron
signals from galaxies and clusters \cite{Thierbach:2002rs}, supposedly
arising from DM annihilation.

The Planck data restrict us to $\Omega h^2 = 0.1199\pm 0.0022$ at the
$1\sigma$ level \cite{Ade:2015xua}, and any viable region in the MSSM
parameter space should obey this limit. However, theoretical
uncertainties, mostly due to higher order corrections, exist in the
computation of the DM annihilation rate, as a result of which
calculated values of the relic density somewhat beyond the above band
is also given the benefit of doubt. Thus regions where the relic
density estimate yields $\Omega h^2$ in the range $0.1199 \pm 0.012$
are taken to be consistent \cite{Harz:2016dql, Klasen:2016qyz,
  Badziak:2017uto}.

In most recent studies, however, the lower bound on $\Omega h^2$ is
not taken as seriously as the upper one \cite{Achterberg:2017emt,
  Caron:2015wda, Bertone:2015tza, Calore:2014nla, Butter:2016tjc}, and
regions that predict under-abundance are taken as allowed. But that
requires additional sources of DM particles, and thus stepping beyond
the MSSM as the new physics scenario at low energy. {\em The present
  work aims to analyse observed data in terms of MSSM in the strict
  sense.}

Regions in the MSSM parameter space have emerged as good fits
for spectra (Case 1) where all sfermions and the gluino are heavy
(above 2 TeV) \cite{Achterberg:2017emt, Caron:2015wda, Bertone:2015tza, Calore:2014nla}.  Another possibility (Case 2) \cite{Achterberg:2017emt,
  Caron:2015wda, Bertone:2015tza, Calore:2014nla, Butter:2016tjc} is to have one light
stop mass eigenstate in the range of $260 - 300$ GeV
\cite{Sirunyan:2017cwe, Aaboud:2017ayj, Aaboud:2017nfd}. An 
elaborate analysis \cite{our_paper} reveals that these two
  cases, for certain ranges of the pseudoscalar Higgs ($m_A$) and
  $\tan\beta$ (the ratio of the vacuum expectation values (vev) of the
  two Higgs doublets) \cite{Aaboud:2016cre, Aaboud:2017sjh}, are the
  fittest, though with differing statistical significance, when one
  demands a minimum neutralino ($\chi^0_1$) relic density. The
  parameters values scanned over in the two cases are summarised in
  Table \ref{Table_benchmark}.

\begin{table}[h]
\centering
\begin{tabular}{|c|c|c|c|c|}

\hline 
Case no & $\tan\beta$ & $M_1, M_2$ & $\mu$ & $m_A$ \\ 
\hline
1  & 20 & $[-1000, 1500]$ & $[-1000, 2000]$ & $[450,  4000]$ \\ 
 
\hline
2  & 50 & $[-1000, 1500]$ & $[-1000, 2000]$ & $[850,  4000]$ \\

\hline
\end{tabular}
\caption{The ranges over which the MSSM parameters have been varied in
  the $\chi^2$-fit for the two benchmark cases which yield the best
  fits to the $\gamma$-ray data. All sfermion masses are above 2 TeV,
  excepting the lighter stop in case 2 whose range is mentioned in the
  text. The parameter $A_t$ is adjusted together with
  $m_{\tilde{t}_1}$ to fix $m_h$ in the range $122 - 128$ GeV and to
  retain a proper electroweak vacuum. All masses are in GeV.}
\label{Table_benchmark}
\end{table}

The $\gamma$-ray flux from the GC is obtained as \cite{Calore:2014xka}:
\begin{equation}  \label{eq:flux1}
\phi(E) = \frac{\left\langle \sigma v \right\rangle}{8 \pi
  m^{2}_{\chi^0_1}} \frac{dN_{\gamma}}{dE}(E) \int_{l.o.s}
\rho^{2}(r(s,\theta)) ~ds ~d\Omega
\end{equation}
where  $\frac{dN_{\gamma}}{dE}$ is the photon energy distribution, and
$\left\langle \sigma v \right\rangle$ is the velocity averaged
total annihilation cross-section of the neutralino DM times the relative 
velocity of the annihilating pair. It is mostly driven by
the $W^+ W^-$ and $b\bar{b}$ channels in Case 1, whereas the
$t\bar{t}$ channel has an important role in Case 2. The integration is
over the line-of-sight variable, $l.o.s$ ($ds$) and the solid angle ($d\Omega$)
subtended at the observation point. $\rho(r)$ is the DM profile in the
GC, and is related to the so-called J-factor by
\begin{equation}
J(\theta) = \int_{l.o.s} \rho^{2}(r(s,\theta)) ~ds
\end{equation}
so that the flux per unit solid angle is given by
\begin{equation}  \label{eq:flux2}
\frac{d\phi}{d\Omega} = \frac{\left\langle \sigma v \right\rangle}{8 \pi
  m^{2}_{\chi^0_1}} \frac{dN_{\gamma}}{dE}(E) J_{av}
\end{equation}
with $J_{av} = \int J(\theta) d\Omega/\int d\Omega$. Here we have used
the $2\sigma$ maximum value of $J_{av}$ according to the
Navarro-Frenk-White (NFW) profile \cite{Bertone:2015tza}. This yields
the most optimistic fit of the GC $\gamma$-ray excess in terms of the
MSSM.

We perform $\chi^2$-fits of the GC spectrum in the range $0.5 - 200$
GeV \cite{Calore:2014xka, TheFermi-LAT:2017vmf}, assuming that all of
the excess arises from annihilation of dark matter particles. We have
used the package micrOMEGAs 4.3.1 for $\langle \sigma v\rangle$
calculation and \cite{Calore:2014xka, GC_matrix} for obtaining the GC
excess data and errors. Scanning is done over $M_1$ and $M_2$ which
are the U(1) and SU(2) gaugino masses, the neutral pseudoscalar mass
($m_A$) and the Higgsino mass parameter $\mu$.  The remaining
superparticle masses are above 2 TeV in Case 1.  For
Case 2, an optimally light stop mass eigenstate at $260 - 300$ GeV is
kept, which facilitates the $t\bar{t}$ annihilation
channel. Consistency with the data from the Large Electron Positron
(LEP) \cite{chargino} and LHC \cite{Sirunyan:2017cwe, Aaboud:2017ayj,
  Aaboud:2017nfd, Aaboud:2016cre, Aaboud:2017sjh}, including the 125
GeV scalar \cite{Aad:2014aba, Khachatryan:2014jba, Allanach:2004rh},
and all other phenomenological constraints are ensured. The most
optimistic value of $\tan\beta$, the ratio of the two Higgs doublet
vev's, has been used in each case.  Parameters such as the
trilinear SUSY-breaking parameters $A_f$, (where f stands for
fermions) are adjusted to reproduce a 125 GeV neutral scalar.

The DM annihilation rate explaining the GC
$\gamma$-ray excess must also be consistent with the flux from other
$\gamma$-ray emitters such as dwarf galaxies. While only upper limits
exist for most of these galaxies \cite{Ackermann:2013yva}, some excess
was claimed for Reticulum II some time ago (in the Pass 7 data)
\cite{Geringer-Sameth:2015lua}, to be replaced by upper limits again
(the Pass 8 data) \cite{Zhao:2017pcz, Drlica-Wagner:2015xua}. It has
been found \cite{our_paper} that the MSSM fits to the GC excess do not change
appreciably whether one is using the Pass 7 excess or the Pass 8 upper
limits. The results reported below are consistent with either.

\begin{figure*}
\centering
  \includegraphics[height=0.25\textwidth, angle=0]{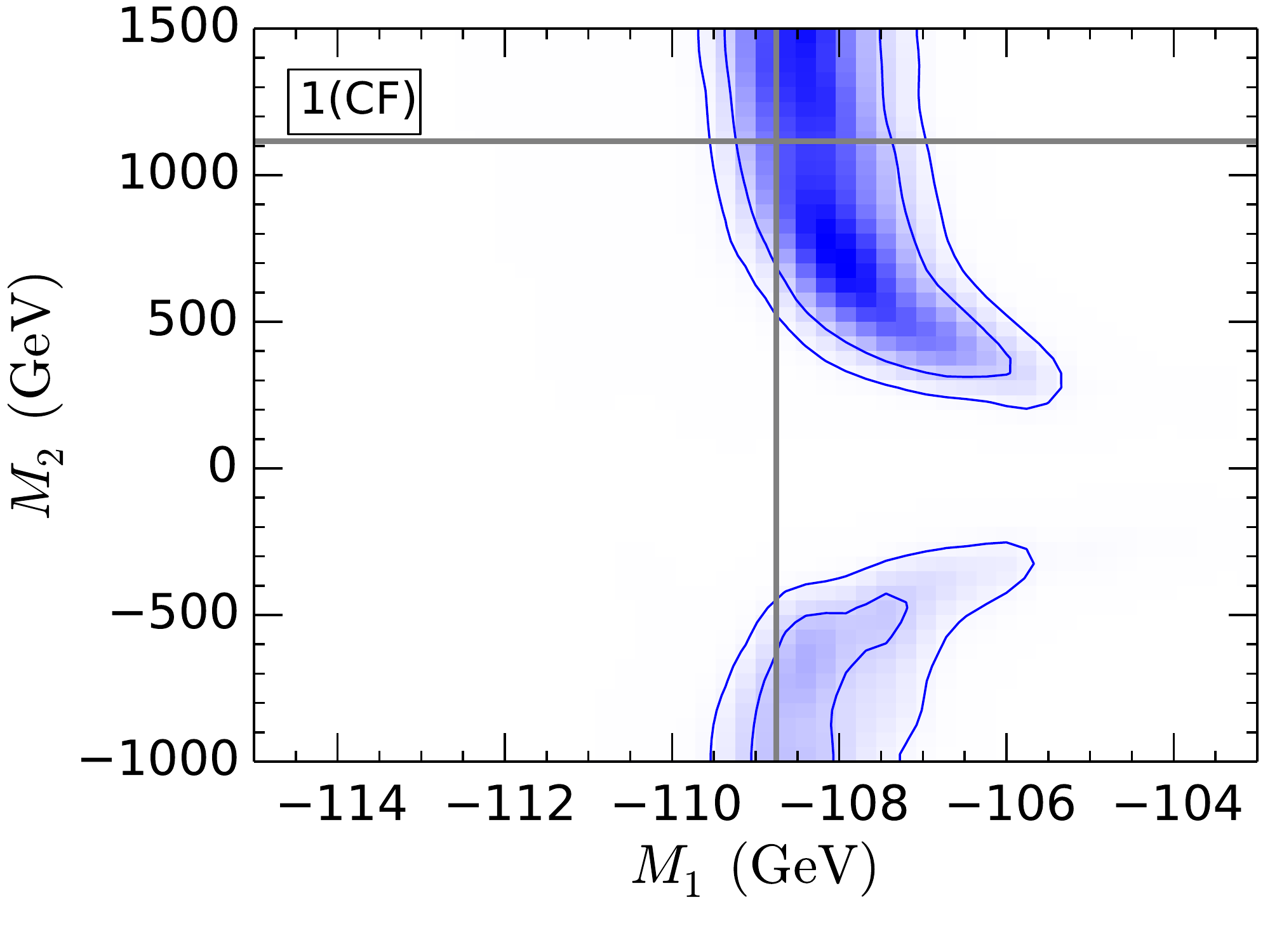}\hspace{18mm}%
  \includegraphics[height=0.25\textwidth, angle=0]{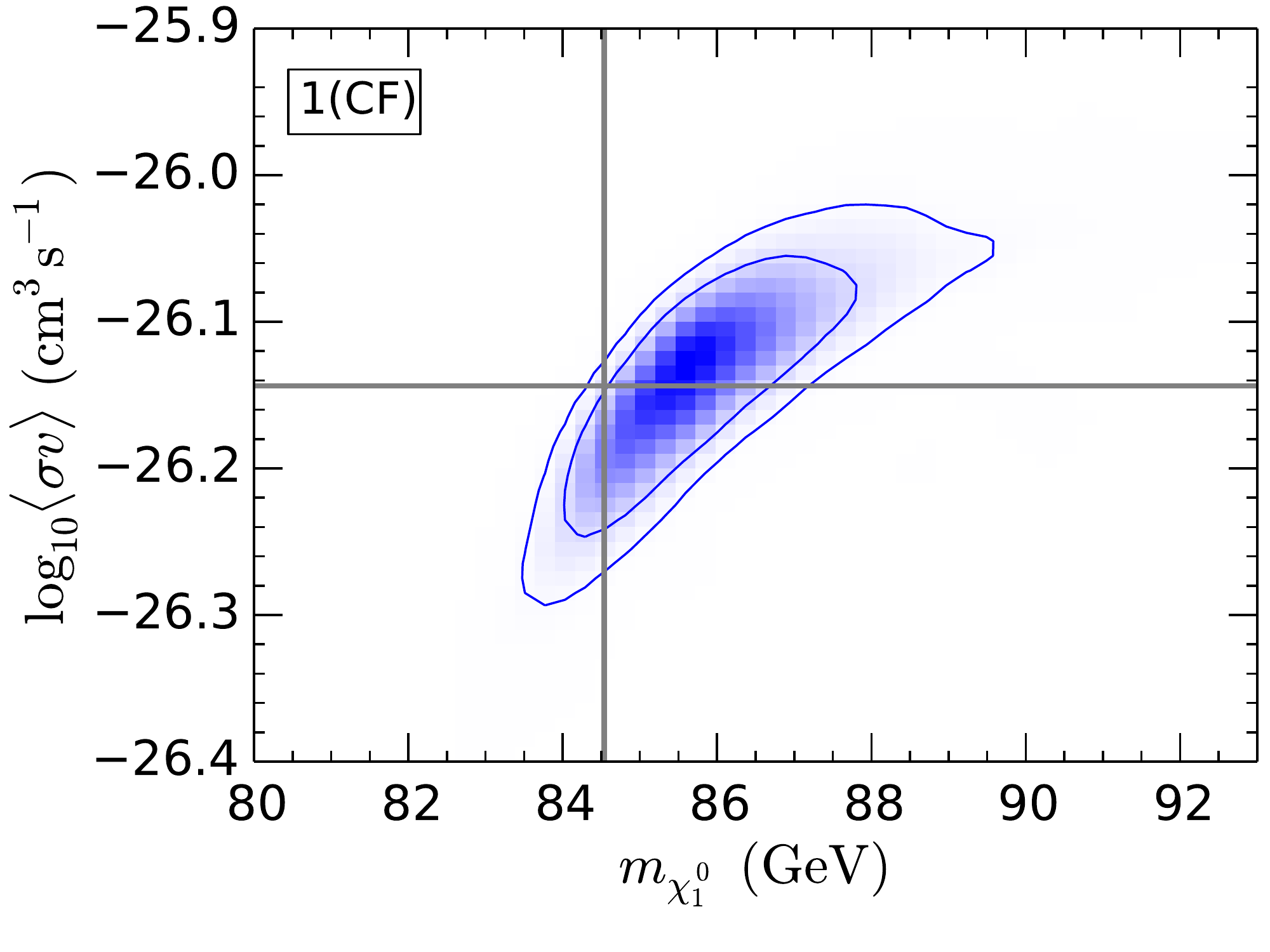}
  \includegraphics[height=0.25\textwidth, angle=0]{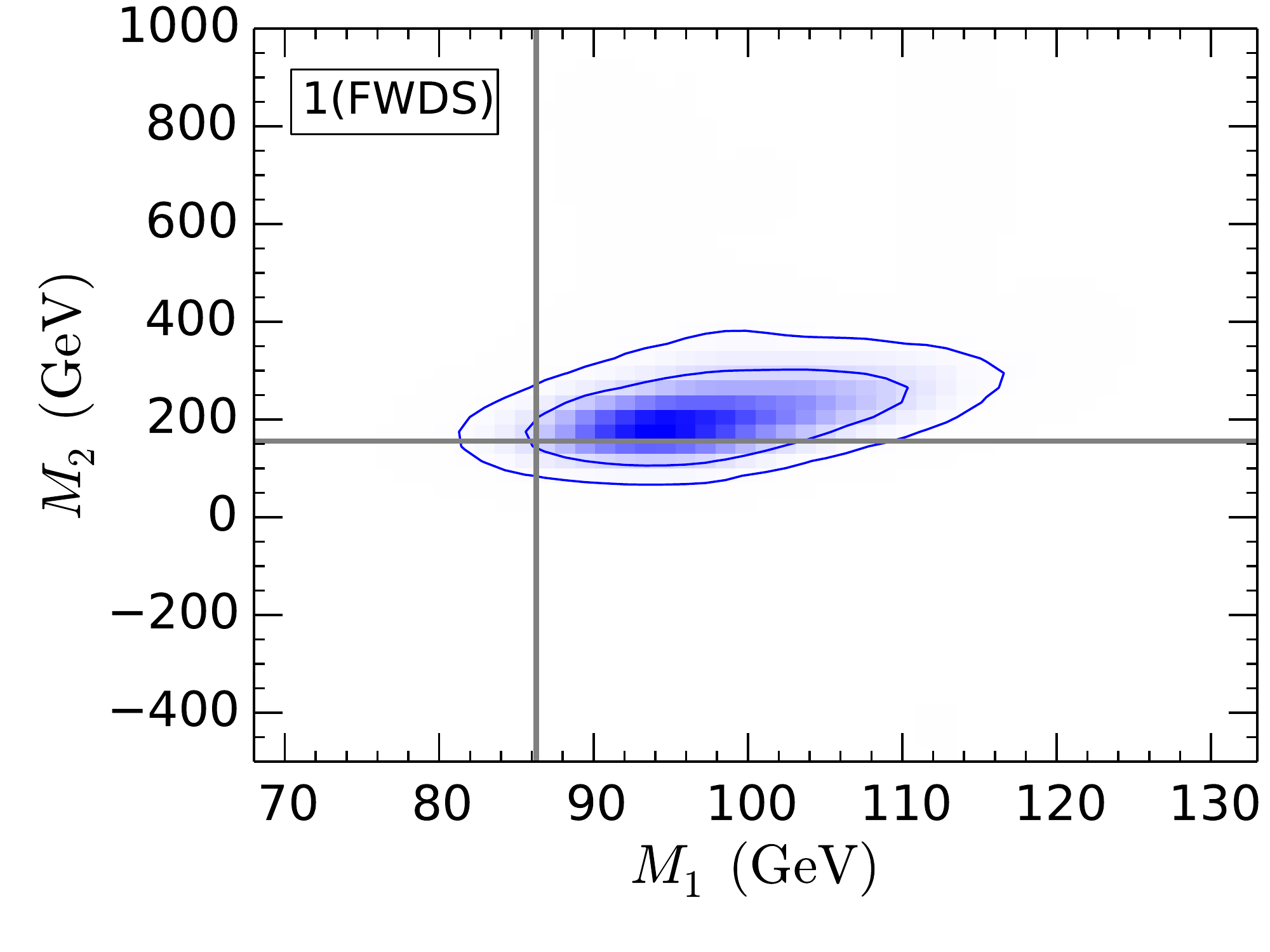}\hspace{18mm}%
  \includegraphics[height=0.25\textwidth, angle=0]{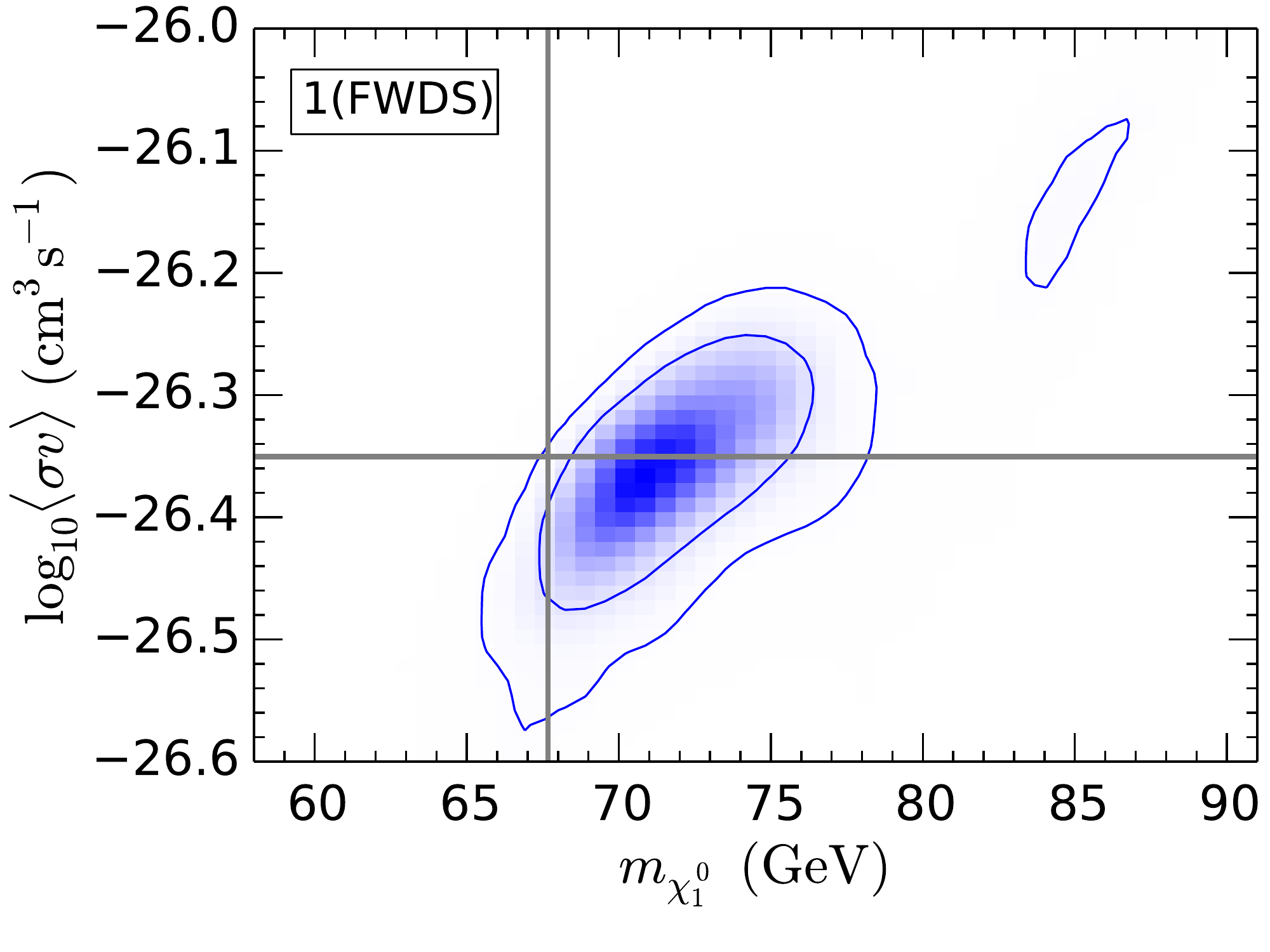}
  \includegraphics[height=0.25\textwidth, angle=0]{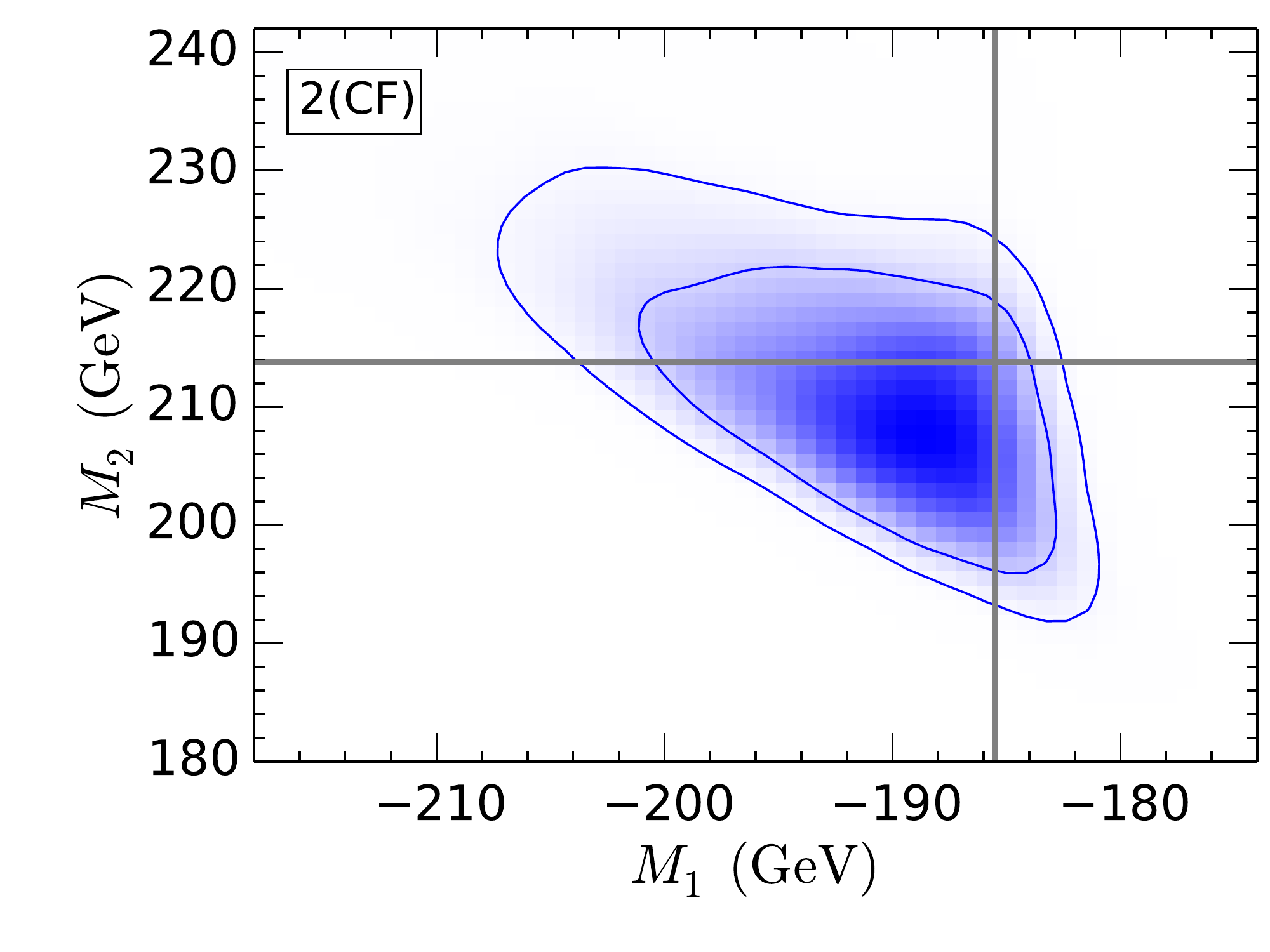}\hspace{18mm}%
  \includegraphics[height=0.25\textwidth, angle=0]{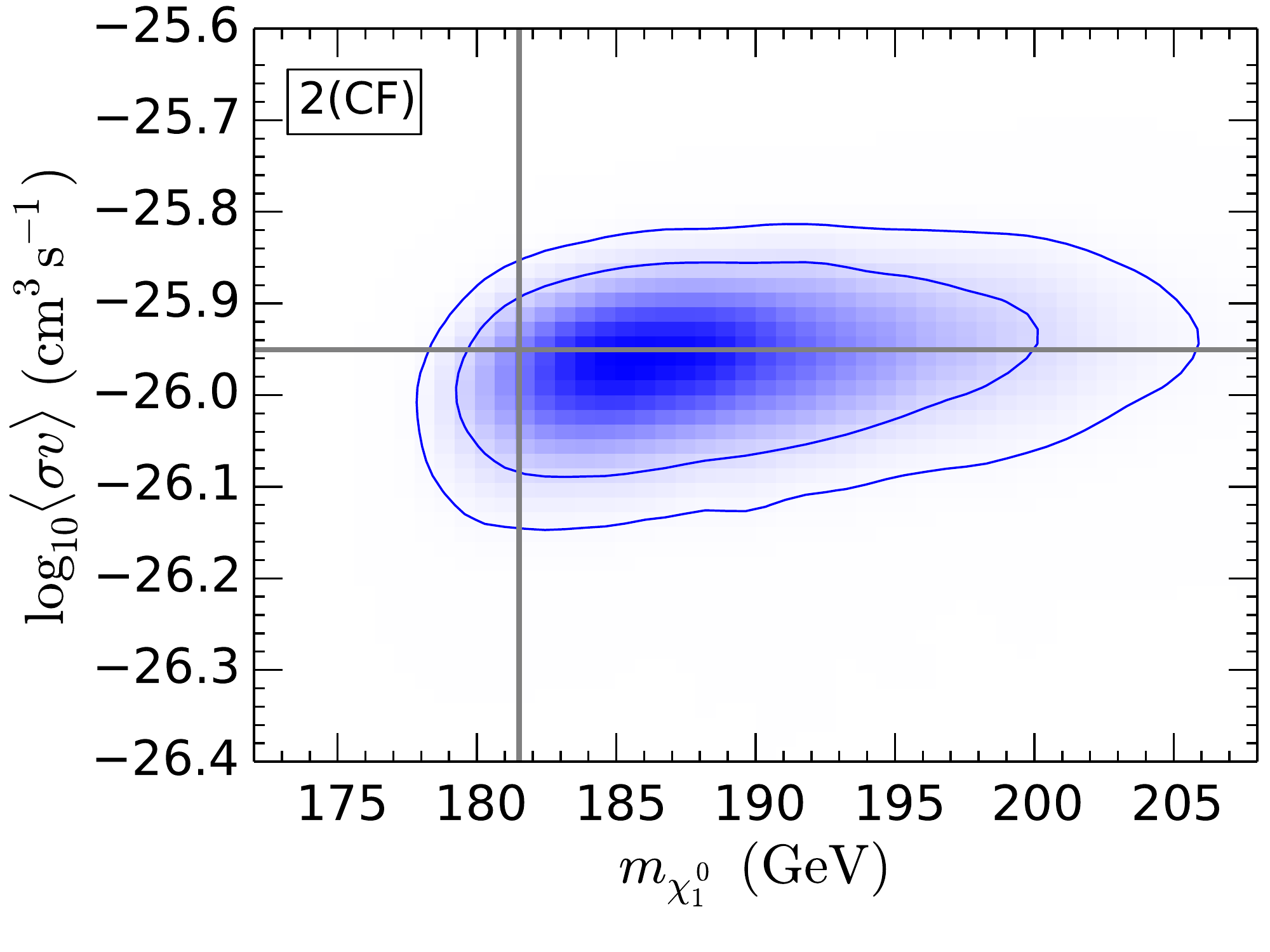}
  \includegraphics[height=0.25\textwidth, angle=0]{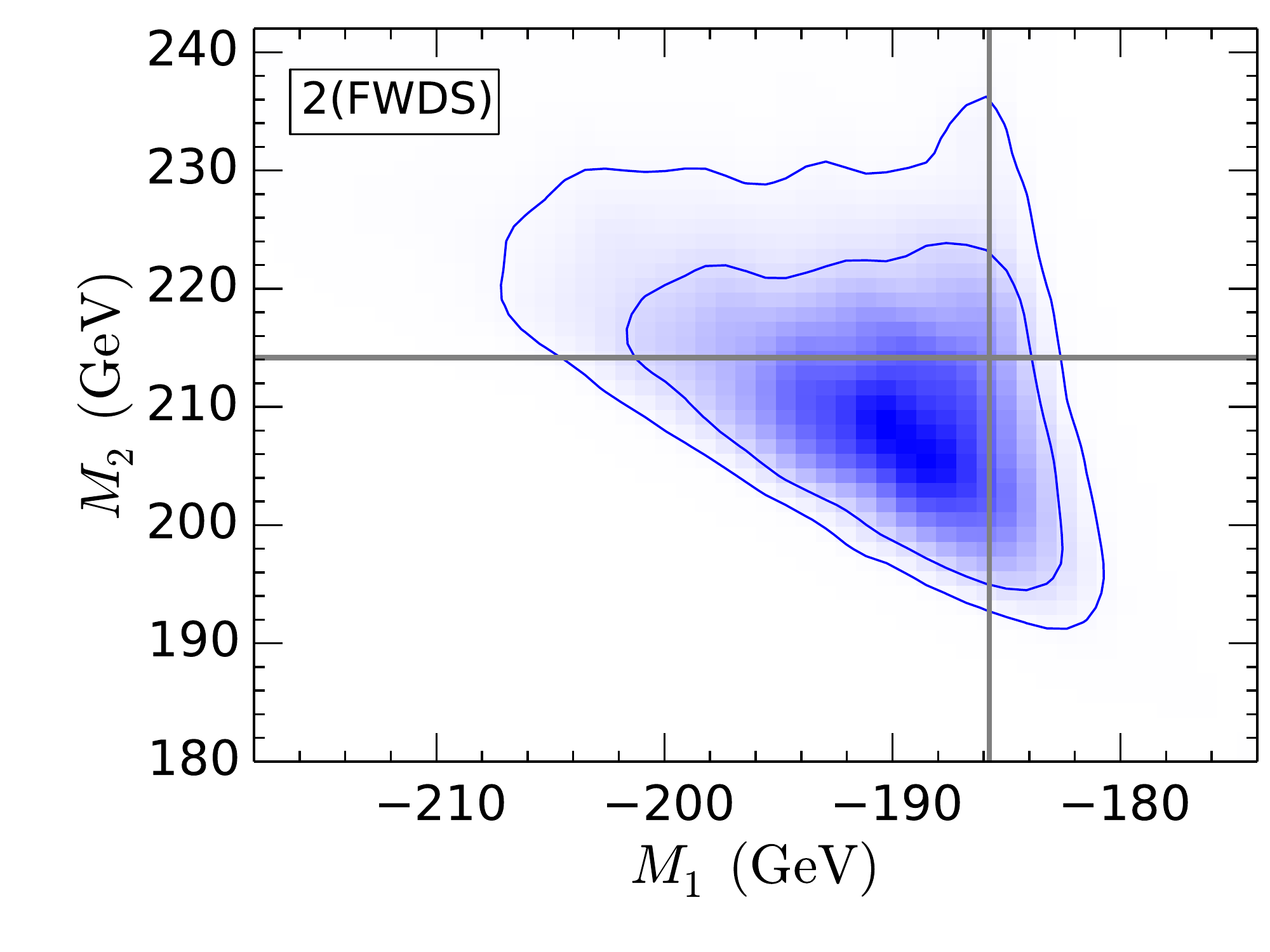}\hspace{18mm}%
  \includegraphics[height=0.25\textwidth, angle=0]{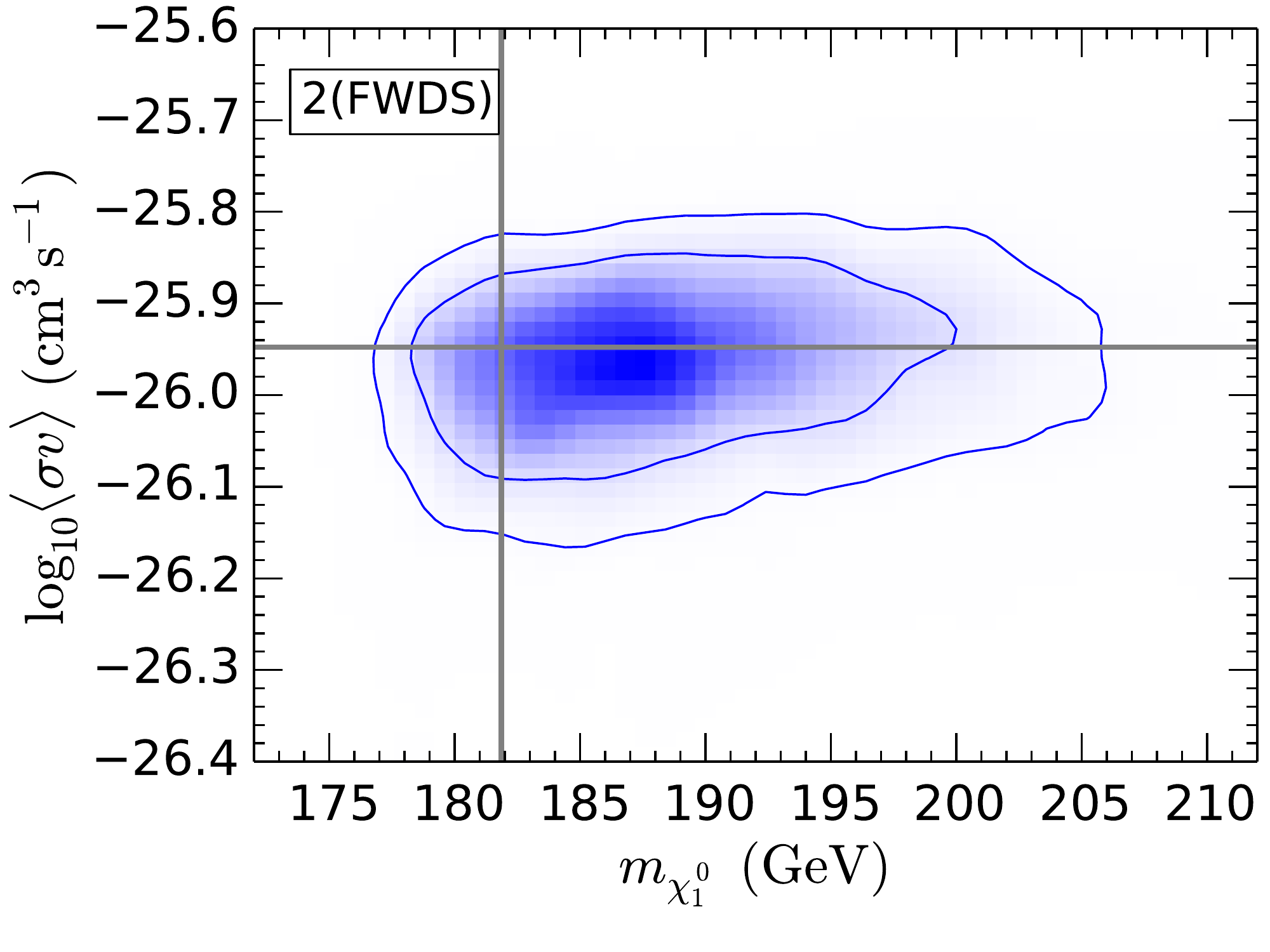}
  \includegraphics[height=0.25\textwidth, angle=0]{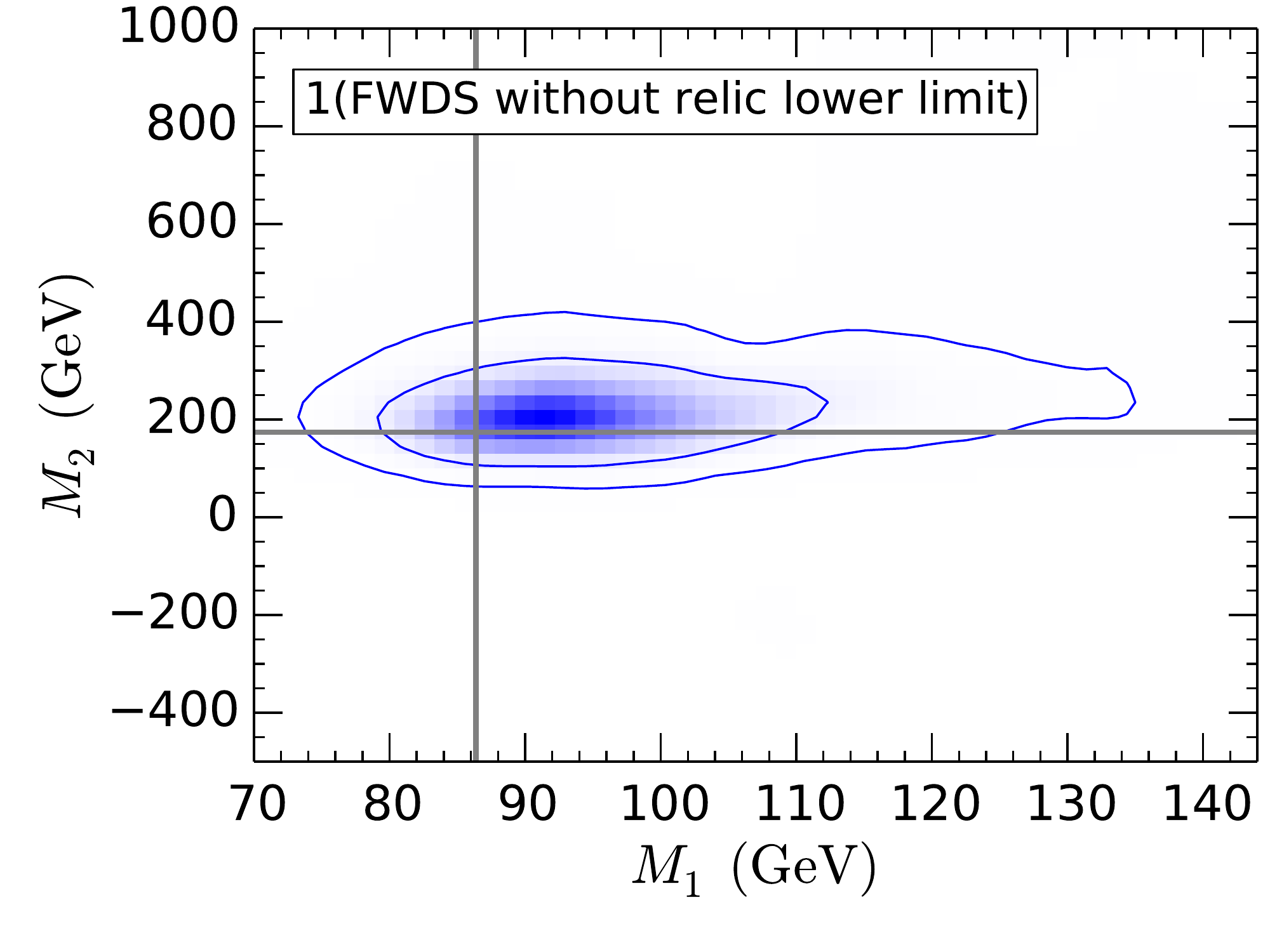}\hspace{18mm}%
  \includegraphics[height=0.25\textwidth, angle=0]{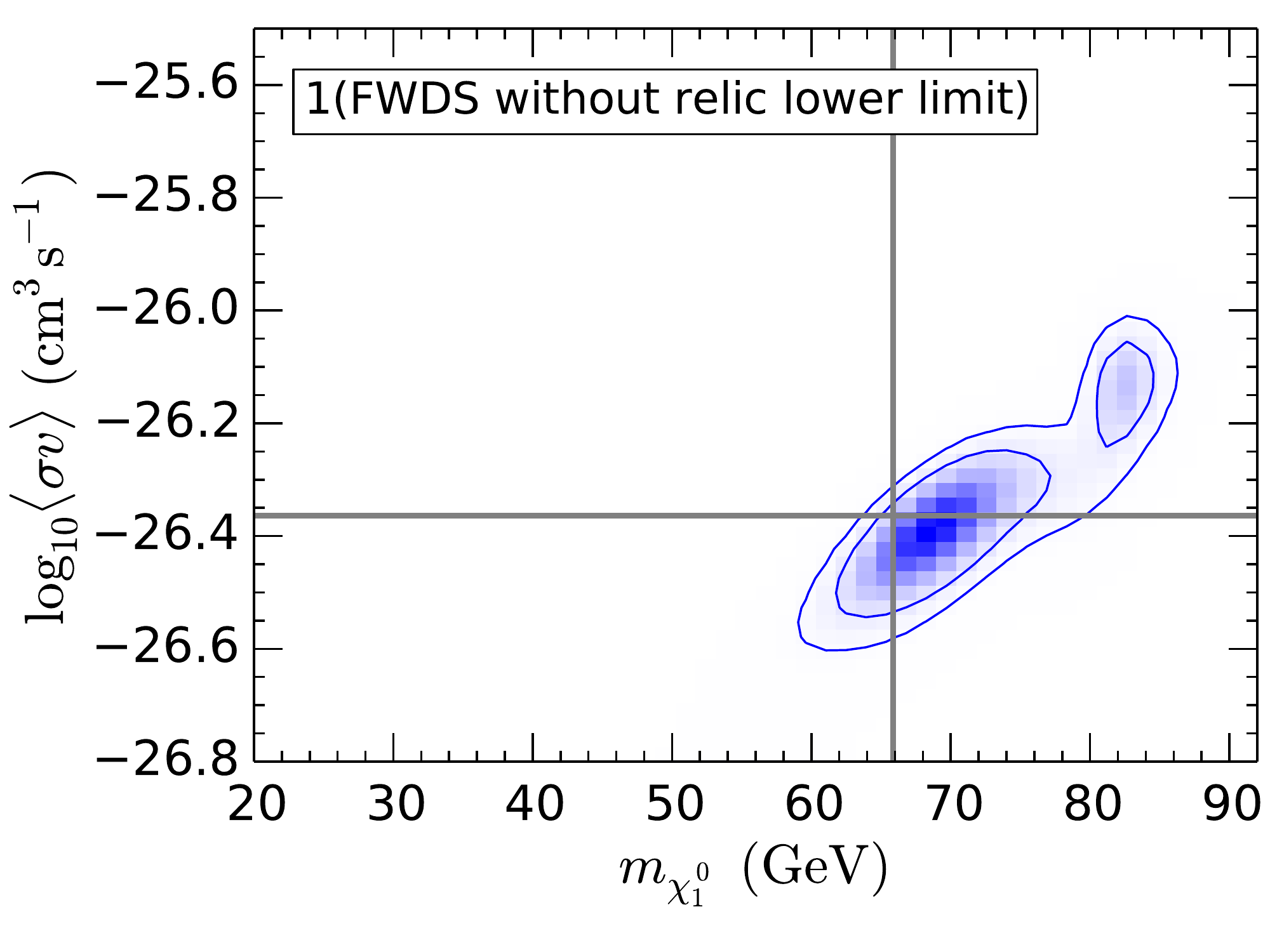}
  
  \caption{$1\sigma$ (inner region) and $2\sigma$ (outer) joint
    constraints on $M_1 - M_2$ (left) and $m_{\chi^0_1} - \langle
    \sigma v \rangle$ (right) along with their best-fit values (solid
    gray lines) for various scenarios mentioned in the text. From top
    to bottom - {\it 1st panel}: 1(CF); {\it 2nd panel}: 1(FWDS); {\it
      3rd panel}: 2(CF); {\it 4th panel}: 2(FWDS); and {\it 5th
      panel}: 1(FWDS without relic lower limit).}

\label{1a_2b}
\end{figure*}

Figure \ref{1a_2b} contains the marginalised $2\sigma$ contours in
various pairs of parameters, fundamental as well as derived, obtained
via $\chi^2$-minimisation, using the Markov Chain Monte Carlo (MCMC)
technique \cite{MCMC}.  They include two kinds of fits: (a) Complete
fits (CF), or fits where all constraints including those from direct
DM search is used to restrict the parameter pace used for the scan,
and (b) Fit without direct search (FWDS), where 
$\chi^2$-minimisation is not biased by the direct search
constraints which are applied only later.

It is obvious from Figure \ref{1a_2b} that, for Case 1, the CF
$2\sigma$ region falls outside the corresponding region for FWDS. In
other words, the $2\sigma$ region in a completely unbiased analysis of
the data gets disallowed when the latest direct search constraint
\cite{Aprile:2017iyp} is imposed.  This because the DM neutralino for
such a spectrum has an appreciable Higgsino component in order to
facilitate annihilation.  This, however, faces severe restriction from
direct search data, and the $2\sigma$ regions in FWDS does not survive
this tug-of-war. The best fit points, barely surviving phenomenological
constraints, are close to the edge of the $2\sigma$ region.

For Case 2, the direct search constraint is found to be much less
restrictive. The $t\bar{t}$ annihilation channel is open here,
requiring less participation of the Higgsino component in $\chi^0_1$
because of a light stop. Such a spectrum is less affected by the direct
search limits. Thus there is a good overlap between the $2\sigma$
regions of FWDS and CF. However, the $t\bar{t}$ channel shifts the
peak of the resulting $\gamma$-ray distribution to energies higher
than what is observed.  This increases the contribution to $\chi^2$
from regions of lower energies \cite{TheFermi-LAT:2015kwa,
  Calore:2014xka}.  Any effort to get a better fit by going to
parameter regions that scale up the distribution fails, because (a)
such regions require a lighter $\chi^0_1$ than what is consistent with
the light stop \cite{Sirunyan:2017cwe, Aaboud:2017ayj,
  Aaboud:2017nfd}, and (b) the annihilation rate becomes so high
that the relic density drops below its lowest admissible value.

\begin{figure}
\begin{center}
\includegraphics[height=0.30\textwidth, angle=0]{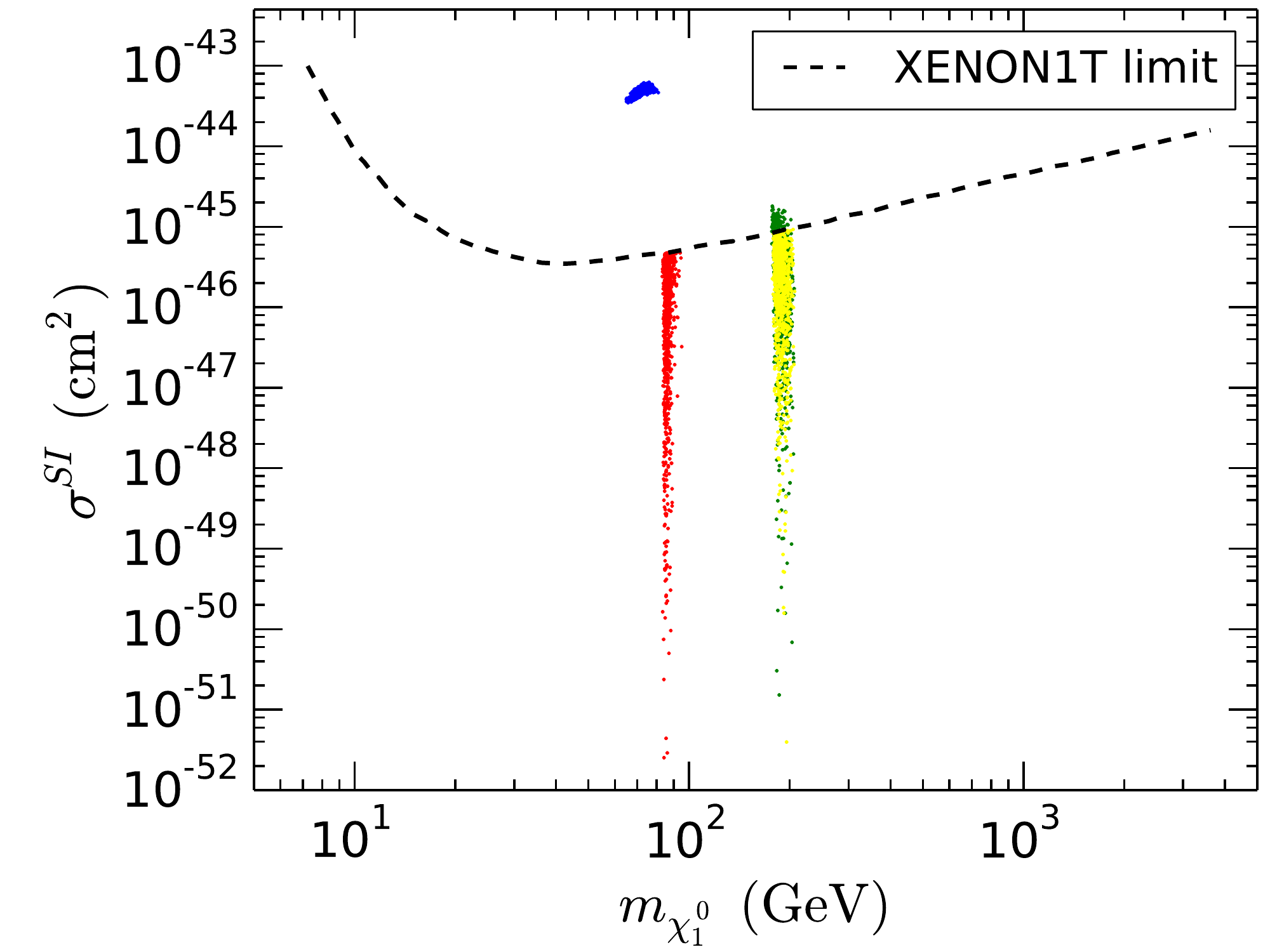}
\caption{95.6\% C.L. region in the $m_{\chi^0_1} - \sigma^{SI}$ plane
  for various scenarios of cases 1 and 2. Red: 1(CF). Blue:
  1(FWDS). Yellow: 2(CF). Green: 2(FWDS). The black line shows the
  upper limit on the spin-independent cross section from XENON1T
  experiment.}
\label{direct_search}
\end{center}	
\end{figure}

All of the above features, especially the effect of direct search limits,
are seen in Figure \ref{direct_search}.
The $2\sigma$ region in Case 1 for FWDS is thus disallowed whereas
that for Case 2 are relatively unaffected. Figure \ref{1a_2b} shows
the result of imposing a lower bound on the MSSM contribution to
the relic density. One finds on comparing the lowest row in Figure 1 
with the topmost one that this
condition plays a role even before doing FWDS. The disallowed regions
correspond to low $m_{\chi^0_1}$ and high $\langle \sigma v \rangle$.  
On the whole, Case 1 remains the best
fit of the GC $\gamma$-ray spectrum, in terms of annihilation rate as
well as the shape of the spectrum \cite{TheFermi-LAT:2015kwa,
  Calore:2014xka}.

In Table \ref{Table_pvalue},  values of
$\chi^2_{min}$ for FWDS as well as CF are presented for both Cases 1
and 2. We also show the corresponding $p$-values which quantify
the probabilities of the MSSM being a good fit for the GC data modulo all
constraints.  It is seen that the $p$-value progressively deteriorates
from Case 1 to Case 2, and from the fit without relic lower limit to 
FWDS and then to CF in Case 1. On the whole,
while Case 1 suffers from the lower limit on the relic density as well
as the direct search constraint, Case 2,  less affected by both,
has an intrinsic disadvantage due to the shape of the
energy distribution of  GC $\gamma$-rays coming from the $t\bar{t}$ channel.

\begin{table}[h]
\centering
\begin{tabular}{|c|c|c|c|}
\hline 
Case no & $\chi^{2}_{min}$/DOF & $p$-value \\ 
\hline
1 (FWDS without relic lower limit) & 44.2/29 & $4 \times 10^{-2}$ \\
1 (FWDS) & 45.1/29 & $3 \times 10^{-2}$ \\
1 (CF) & 56.2/29 & $2 \times 10^{-3}$\\ 
 
\hline
2 (FWDS) & 66.1/29 & $1 \times 10^{-4}$  \\ 
2 (CF) & 66.2/29 & $1 \times 10^{-4}$ \\ 
 
\hline
\end{tabular}
\caption{Comparison of the quality of fitting between different types of 
analysis for Cases 1 and 2. DOF stands for degrees of freedom whose value
is 29 when the Reticulum II excess bins are counted.}
\label{Table_pvalue}
\end{table}

We have already said that our benchmarks for Cases 1 and 2
represent rather faithfully the MSSM spectra that
yield the best possible fits. The other important candidate scenario
is one with a light stau close (within 4\%) to $\chi^0_1$, where
co-annihilation causes freeze-out of the latter.  It however leads to
a fit worse than the two cases reported above \cite{our_paper}.  This
is because the optimum region for this in the MSSM parameter space
corresponds to a somewhat heavier $\chi^0_1 - \tilde{\tau}_1$ pair,
which again shifts the GC $\gamma$-ray distribution peak to
unacceptably high values. The situation with both the $\tilde{\tau}_1$
and the $\tilde{t}_1$ light is also worse off for similar reasons.

Finally, we contrast this with a comparable situation in the context
of Higgs boson search. Before the LHC data came up with anything
decisive, there was a lower bound on the Higgs boson mass of about 114
GeV from LEP \cite{Baak:2011ze, Flacher:2008zq}. Similarly, the
Tevatron data strongly disfavoured the mass range 158 GeV $< m_h <$
172 GeV \cite{Baak:2011ze}. On the other hand, the best fit value of
$m_h$ from indirect data including those from precision electroweak
measurements was less than 100 GeV, ruled out by direct searches.
However, the $2\sigma$ region in $\chi^2$ extended upto $m_h \approx
160 - 170$ GeV. Though the LEP and Tevatron limits removed regions
with the $2\sigma$ range of such a `standard fit', it left out the
value 125 GeV at the edge of the $1\sigma$ band, and that is where the
particle was finally discovered at the LHC. When instead a `complete
fit' was performed in July, 2012, assigning large values to $\chi^2$
in the bands disallowed (disfavoured) by LEP (Tevatron), and duly
weighing in the LHC data analysed till then, the best fit point was
close to 120 GeV. Also, $\chi^2_{min}$/DOF converged from 16.6/13 in
the `standard fit' to 17.8/14 in the `complete fit' \cite{Baak:2011ze,
  Flacher:2008zq}. The $p$-value, too, increased convergently from
0.21 to 0.23. This showed that the successive imposition of
constraints was improving the search, taking one closer to actual
discovery. Thus it became clear that {\em the Glashow-Salam-Weinberg
  theory is a good fit for Higgs-related data.} In contrast, the
  above analysis on GC $\gamma$-ray data, whose results are adumbrated
  in Table \ref{Table_pvalue} and Figures \ref{1a_2b} and
  \ref{direct_search}, indicate that {\em the MSSM is not a good fit
  the available astrophysical observations.} The requirement that the
  entire relic density must be explained by the MSSM (without which it
  is not strictly `minimal'), as also direct search constraints,
  has a role in this conclusion. This suggestion loses ground only if
  (a) the $\gamma$-ray excess from GC is due to some yet unknown
  astrophysical effect, or (b) a large body evidence from alternative
  astronomical observations render the GC excess insignificant.

 {\bf Acknowledgements:}\\
 We thank A. Biswas,  U. Chattopadhyay, S. Mondal and N. Paul 
 for helpful comments. The work of AK and BM was partially supported
 by funding available from the Department of Atomic Energy, Government
 of India, for the Regional Centre for Accelerator-based Particle
 Physics (RECAPP), Harish-Chandra Research Institute. TRC acknowledges
 the hospitality of RECAPP while the project was being formulated,
 while AK and BM thank the National Centre for Radio Astrophysics,
 Pune, for hospitality while work was in progress.


%


\begin{thebibliography}{99}


\bibitem{TheFermi-LAT:2015kwa} 
  M.~Ajello {\it et al.} [Fermi-LAT Collaboration],
  Astrophys.\ J.\  {\bf 819}, 44 (2016)
  [arXiv:1511.02938 [astro-ph.HE]].





\bibitem{Calore:2014xka} 
  F.~Calore, I.~Cholis and C.~Weniger,
  JCAP {\bf 1503}, 038 (2015)
  [arXiv:1409.0042 [astro-ph.CO]].
  
  


\bibitem{TheFermi-LAT:2017vmf} 
  M.~Ackermann {\it et al.} [Fermi-LAT Collaboration],
  Astrophys.\ J.\  {\bf 840}, 43 (2017)
  [arXiv:1704.03910 [astro-ph.HE]].
  


\bibitem{Geringer-Sameth:2015lua} 
  A.~Geringer-Sameth, M.~G.~Walker, S.~M.~Koushiappas, S.~E.~Koposov, V.~Belokurov, G.~Torrealba and N.~W.~Evans,
  Phys.\ Rev.\ Lett.\  {\bf 115}, 081101 (2015)
  [arXiv:1503.02320 [astro-ph.HE]].
  
  
  



\bibitem{Aprile:2017iyp} 
  E.~Aprile {\it et al.} [XENON Collaboration],
  Phys.\ Rev.\ Lett.\  {\bf 119}, 181301 (2017)
  [arXiv:1705.06655 [astro-ph.CO]].
  
  
  
\bibitem{PhysRevLett.118.251302}
D. S. Akerib {\it et al.}. [LUX Collaboration], 
Phys. Rev. Lett. {\bf 118}, 251302, 2017




  



\bibitem{Baak:2011ze} 
  M.~Baak, M.~Goebel, J.~Haller, A.~Hoecker, D.~Ludwig, K.~Moenig, M.~Schott and J.~Stelzer,
  Eur.\ Phys.\ J.\ C {\bf 72}, 2003 (2012)
  [arXiv:1107.0975 [hep-ph]].
  
  
  
  
\bibitem{Flacher:2008zq} 
  H.~Flacher, M.~Goebel, J.~Haller, A.~Hocker, K.~Monig and J.~Stelzer,
  Eur.\ Phys.\ J.\ C {\bf 60}, 543 (2009)
  Erratum: [Eur.\ Phys.\ J.\ C {\bf 71}, 1718 (2011)]
  [arXiv:0811.0009 [hep-ph]].
  








\bibitem{Sirunyan:2017cwe} 
  A.~M.~Sirunyan {\it et al.} [CMS Collaboration],
  Phys.\ Rev.\ D {\bf 96}, 032003 (2017)
  [arXiv:1704.07781 [hep-ex]].




\bibitem{Aaboud:2017bac} 
  M.~Aaboud {\it et al.} [ATLAS Collaboration],
  arXiv:1708.08232 [hep-ex].






  

  
  
  
 
 
\bibitem{Zhao:2017pcz} 
  Y.~Zhao, X.~J.~Bi, P.~F.~Yin and X.~M.~Zhang,
  arXiv:1702.05266 [astro-ph.HE].





\bibitem{Drlica-Wagner:2015xua} 
  A.~Drlica-Wagner {\it et al.} [Fermi-LAT and DES Collaborations],
  Astrophys.\ J.\  {\bf 809}, L4 (2015)
  [arXiv:1503.02632 [astro-ph.HE]].
  
  
  
\bibitem{Ackermann:2017nya} 
  M.~Ackermann {\it et al.} [Fermi-LAT Collaboration],
  Astrophys.\ J.\  {\bf 836}, 208 (2017)
  [arXiv:1702.08602 [astro-ph.HE]].
  
            



\bibitem{Thierbach:2002rs} 
  M.~Thierbach, U.~Klein and R.~Wielebinski,
  Astron.\ Astrophys.\  {\bf 397}, 53 (2003)
  [astro-ph/0210147].
  
  
  
  
\bibitem{Ade:2015xua} 
  P.~A.~R.~Ade {\it et al.} [Planck Collaboration],
  Astron.\ Astrophys.\  {\bf 594}, A13 (2016)
  [arXiv:1502.01589 [astro-ph.CO]].
  





\bibitem{Harz:2016dql} 
  J.~Harz, B.~Herrmann, M.~Klasen, K.~Kovarik and P.~Steppeler,
  Phys.\ Rev.\ D {\bf 93}, 114023 (2016)
  [arXiv:1602.08103 [hep-ph]]. 
  
  
  
  
  
\bibitem{Klasen:2016qyz} 
  M.~Klasen, K.~Kovarik and P.~Steppeler,
  Phys.\ Rev.\ D {\bf 94}, 095002 (2016)
  [arXiv:1607.06396 [hep-ph]].
  
  
  

\bibitem{Badziak:2017uto} 
  M.~Badziak, M.~Olechowski and P.~Szczerbiak,
  JHEP {\bf 1707}, 050 (2017)
  [arXiv:1705.00227 [hep-ph]].




\bibitem{Achterberg:2017emt} 
  A.~Achterberg, M.~van Beekveld, S.~Caron, G.~A.~Gómez-Vargas, L.~Hendriks and R.~Ruiz de Austri,
  arXiv:1709.10429 [hep-ph].
  
  
  
  
  
  
\bibitem{Caron:2015wda} 
  A.~Achterberg, S.~Amoroso, S.~Caron, L.~Hendriks, R.~Ruiz de Austri and C.~Weniger,
  JCAP {\bf 1508}, 006 (2015)
  [arXiv:1502.05703 [hep-ph]].
  
  
  
  
  
  
\bibitem{Bertone:2015tza} 
  G.~Bertone, F.~Calore, S.~Caron, R.~Ruiz, J.~S.~Kim, R.~Trotta and C.~Weniger,
  JCAP {\bf 1604}, 037 (2016)
  [arXiv:1507.07008 [hep-ph]].
  
  
  
  
  
\bibitem{Calore:2014nla} 
  F.~Calore, I.~Cholis, C.~McCabe and C.~Weniger,
  Phys.\ Rev.\ D {\bf 91}, 063003 (2015)
  [arXiv:1411.4647 [hep-ph]].
  
  
  
  
\bibitem{Butter:2016tjc} 
  A.~Butter, S.~Murgia, T.~Plehn and T.~M.~P.~Tait,
  Phys.\ Rev.\ D {\bf 96}, 035036 (2017)
  [arXiv:1612.07115 [hep-ph]].





  
  


\bibitem{Aaboud:2017ayj} 
  M.~Aaboud {\it et al.} [ATLAS Collaboration],
  arXiv:1709.04183 [hep-ex].



\bibitem{Aaboud:2017nfd} 
  M.~Aaboud {\it et al.} [ATLAS Collaboration],
  arXiv:1708.03247 [hep-ex].







\bibitem{our_paper}
A. Kar, S. Mitra, B. Mukhopadhyaya, T. Roy Choudhury (in preparation)




  
  
  
\bibitem{Aaboud:2016cre} 
  M.~Aaboud {\it et al.} [ATLAS Collaboration],
  Eur.\ Phys.\ J.\ C {\bf 76}, 585 (2016)
  [arXiv:1608.00890 [hep-ex]].





\bibitem{Aaboud:2017sjh} 
  M.~Aaboud {\it et al.} [ATLAS Collaboration],
  arXiv:1709.07242 [hep-ex].
  
  
  
  

  
  
  
\bibitem{GC_matrix}
$http://christophweniger.com/?page_{-}id=248$ 





\bibitem{chargino} 
C. Patrignani et al. (Particle Data Group), Chin. Phys. C, {\bf 40}, 100001 (2016) and 2017 update



  
  
  
  
  
  
\bibitem{Aad:2014aba} 
  G.~Aad {\it et al.} [ATLAS Collaboration],
  Phys.\ Rev.\ D {\bf 90}, 052004 (2014)
  [arXiv:1406.3827 [hep-ex]].




\bibitem{Khachatryan:2014jba} 
  V.~Khachatryan {\it et al.} [CMS Collaboration],
  Eur.\ Phys.\ J.\ C {\bf 75}, 212 (2015)
  [arXiv:1412.8662 [hep-ex]].



\bibitem{Allanach:2004rh} 
  B.~C.~Allanach, A.~Djouadi, J.~L.~Kneur, W.~Porod and P.~Slavich,
  JHEP {\bf 0409}, 044 (2004)
  [hep-ph/0406166].
  
  
  
  
\bibitem{Ackermann:2013yva} 
  M.~Ackermann {\it et al.} [Fermi-LAT Collaboration],
  Phys.\ Rev.\ D {\bf 89}, 042001 (2014)
  [arXiv:1310.0828 [astro-ph.HE]].
  
  
  
\bibitem{MCMC}
Daniel Foreman-Mackey {\it et al.}, arXiv:1202.3665 [astro-ph.IM]



  


\end{thebibliography}

\end{document}